\documentclass[smallabstract,smallcaptions]{dccpaper}
\usepackage{longtable}
\usepackage{epsfig}
\usepackage{amsmath}
\usepackage{amssymb}
\usepackage{color}
\usepackage{url}
\usepackage{xspace}
\usepackage{multirow}
\usepackage{threeparttable} 
\usepackage{booktabs}
\usepackage[hidelinks]{hyperref}
\usepackage{threeparttable} 
\usepackage{array}
\usepackage{bigstrut}
\usepackage{adjustbox}
\usepackage{colortbl}
\usepackage{makecell}
\usepackage{amssymb}
\usepackage{xcolor}    
\usepackage{colortbl}
\usepackage{subcaption}
\usepackage{graphicx}
\newlength{\figurewidth}
\newlength{\smallfigurewidth}

\newcommand{\ie}{{\emph{i.e.}}\xspace}
\newcommand{\eg}{{\emph{e.g.}}\xspace}

\begin{document}

\title
{\large
\textbf{Voxel-GS: Quantized Scaffold Gaussian Splatting Compression with Run-Length Coding}
}

\author{%
Chunyang Fu$^{1}$, Xiangrui Liu$^{1}$, Shiqi Wang$^{1}$ and Zhu Li$^{2}$\\[0.5em]
{\small\begin{minipage}{\linewidth}\begin{center}
\begin{tabular}{ccc}
$^{1}$City University of Hong Kong & \hspace*{0.5in} & $^{2}$University of Missouri-Kansas City \\
Hong Kong SAR, China &&  Kansas City, USA \\
\url{chunyang.fu@my.cityu.edu.hk}&&\url{shiqwang@cityu.edu.hk}\\
\url{xiangrliu3-c@my.cityu.edu.hk}&&\url{zhu.li@ieee.org}
\end{tabular}
\end{center}\end{minipage}}
\thanks{The research was partially supported by the RGC General Research Fund 11200323,  NSFC/RGC JRS Project N\_CityU198/24, and Hong Kong Innovation and Technology Fund GHP/044/21SZ, and PRP/036/24FX.}
}

\maketitle
\thispagestyle{empty}

\begin{abstract}
Substantial Gaussian splatting format point clouds require effective compression. In this paper, we propose Voxel-GS, a simple yet highly effective framework that departs from the complex neural entropy models of prior work, instead achieving competitive performance using only a lightweight rate proxy and run-length coding. Specifically, we employ a differentiable quantization to discretize the Gaussian attributes of Scaffold-GS. Subsequently, a Laplacian-based rate proxy is devised to impose an entropy constraint, guiding the generation of high-fidelity and compact reconstructions. Finally, this integer-type Gaussian point cloud is compressed losslessly using Octree and run-length coding. Experiments validate that the proposed rate proxy accurately estimates the bitrate of run-length coding, enabling Voxel-GS to eliminate redundancy and optimize for a more compact representation. Consequently, our method achieves a remarkable compression ratio with significantly faster coding speeds than prior art. The code is available at \href{https://github.com/zb12138/VoxelGS}{https://github.com/zb12138/VoxelGS}.
\end{abstract}

\Section{Introduction}
3D Gaussian Splatting (3DGS)~\cite{kerbl20233d} demonstrated extraordinary abilities in the field of three-dimensional modeling. However, its compression still requires further research, which has attracted considerable attention from both industry and academia. In most existing compression works~\cite{chen2024hac,HAC_Plus,liu2024compgs}, learning-based entropy models (\eg, factorized entropy bottleneck~\cite{balle2017end}) are part of a Gaussian splatting reconstruction network. This joint optimization of reconstruction and compression directly learns multiview images and generates bitstreams, with Gaussian point clouds as a byproduct. Considering this, MPEG proposed the Gaussian splat coding requirements draft~\cite{mpeggs_coding} at the recent 152nd meeting, which considers two paradigms for potential standardization of GS compression schemes:
\begin{itemize}
    \item A-3DGS: ``Alternative'' 3DGS representation that can be trained based on ground-truth images/videos. This scheme is consistent with most current 3DGS compression methods~\cite{chen2024hac,HAC_Plus,liu2024compgs}. They usually incorporate a GS network with a learning-based entropy model for joint learning, which uses a set of multiview images as input and directly generates the bitstream. 
    \item I-3DGS: ``INRIA''~\cite{kerbl20233d} format coding, which involves trained GS data or ``other'' point cloud-based representation. Since 3D GS point clouds have the same format as standard point clouds, it may be reasonable to extend existing point cloud coders (\eg, G-PCC) to support 3D GS compression. 
\end{itemize}  

Although A-3DGS compression methods can achieve a high rate-distortion performance, they primarily face the following issues. 1) It is necessary to save the parameters of the deep entropy model and transmit the neural network for each scene, but when the GS scene is relatively small, the bitrate budget for the neural network is limited. 2) Since reconstruction and compression are optimized together, the content of the reconstruction is highly correlated with the entropy model, making it difficult to reuse or optimize the coder separately based on the existing reconstruction results, which is the most time-consuming part of the entire system. 
In I-3DGS compression, the preceding reconstruction network can be designated as the naive 3DGS~\cite{kerbl20233d}, but this raises some issues. 1) The model's output is floating-point data, which requires a lot of bits for storage to ensure rendering fidelity, which is not favorable for compression. 2) There is no entropy constraint during the reconstruction process, leading to output point clouds that are redundant and not compact enough. 

To address these issues, we discuss a new preceding 3DGS network to generate GS point clouds for I-3DGS in this paper, whose goal is to obtain an explicit, compact GS point cloud during the reconstruction process and then compress it using downstream coders. The proposed 3DGS network for I-3DGS is based on one of the most popular methods, Scaffold-GS~\cite{lu2024scaffold}, and has undergone very minor modifications. Furthermore, we assumed that run-length coding is used as the downstream coder, which is a fundamental and widely used entropy coding method, to make the generated point clouds compatible with a wider range of other downstream codecs. We believe that the proposed scheme in this paper contributes to the standardization of Gaussian point cloud compression.

\Section{Related Works}
The distribution in 3D Gaussian splatting is essentially sparse and disordered. Due to redundancy among Gaussian primitives, they can be pruned by using masks, gradient thresholds, view-dependent metrics, and other mechanisms for significance evaluation. Additionally, the features of Gaussian primitives, such as spherical harmonic (SH) coefficients, opacity, and scales, can be adjusted through quantization, codebooks, and entropy constraints to reduce the uncertainty of the information. Moreover, early exploration works, such as Compressed3D~\cite{niedermayr2024compressed} and Compact3DGS~\cite{lee2024compact}, have pruned GS parameters using mask, codebook, and quantization methods to achieve a compact representation. Scaffold-GS~\cite{lu2024scaffold} uses anchor points to derive Gaussians through multilayer networks, achieving a more compact representation of GS and reducing parameters while improving fidelity. LightGaussian~\cite{LightGaussian} identifies Gaussians with minimal global significance in scene reconstruction and applies a pruning and recovery process to reduce redundancy while preserving visual quality. Although these methods for compact representation are somewhat similar to the first objective discussed in this paper, which is to generate compact representations, they do not utilize explicit entropy constraints, which significantly enhance the compactness of the GS representation in our method.

Subsequent methods, CompGS~\cite{liu2024compgs}, ContextGS~\cite{wang2024contextgs}, HAC~\cite{chen2024hac}, and HAC++~\cite{HAC_Plus}, have gradually shifted towards joint reconstruction and compression, introducing learnable entropy models in GS networks and enhancing context-aware designs to achieve better rate-distortion performance. However, these schemes primarily face two issues: first, the need to store entropy model parameters, which can become a significant overhead in small scenes; second, their lack of flexibility, as the compressor cannot be used or optimized independently of the reconstruction process. Furthermore, FCGS~\cite{chen2024fast} utilizes pretrained Gaussian point clouds for offline learning of entropy models; however, the achieved compression rate is not particularly high. The main reasons might be that the pre-trained Gaussian point clouds are not compact, and the quantization noise introduced in the compression stage deteriorates the rendering quality. The proposed scheme in this paper will be very helpful in alleviating this problem, specifically by generating compact expressions through Laplacian entropy constraints and limiting the output Gaussian point cloud to integers to avoid the noise impact introduced by subsequent compressors. HGSC~\cite{HGSC} and HyBridGS~\cite{yanghybridgs} are recent works that discuss standard I-3DGS compression methods; however, they utilize RAHT as the point cloud compression scheme, which is a lossy compression mode that introduces noise, potentially deteriorating the rendering quality.
\Section{Method}
\begin{figure}[thb]
    \centering
    \includegraphics[width=0.98\linewidth]{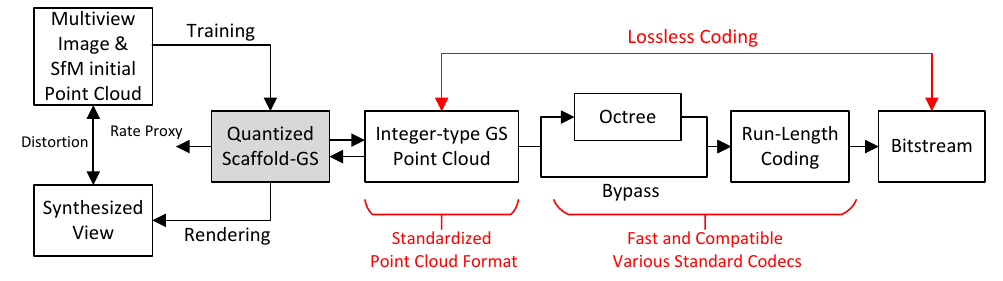}
    \caption{Overview of the proposed Gaussian splatting compression methods.}
    \label{Fig1}
\end{figure}
\noindent\textbf{Overview} The overview of Voxel-GS is shown in Fig.~\ref{Fig1}. Unlike most existing end-to-end Gaussian splatting compression methods, Voxel-GS separates the reconstruction and compression processes. After training of Voxel-GS, a compact, integer-type Gaussian point cloud is obtained, and can be compressed by octree and run-length coding, or other potential codecs. This structure has the following advantages: 1) Separating reconstruction and compression, enabling further research and improving offline Gaussian point cloud codecs. 2)
Losslessly compressing the integer-type features ensures that the compression process does not introduce new distortion, preventing the deterioration of rendering quality. 3) The output Gaussian point clouds can be edited and the compression process is very fast, allowing quick content distribution without undergoing a lengthy reconstruction process. 4) Provides a method for standardization research on I-3DGS compression.

\textbf{Voxelized 3D-GS Representation} Scaffold-GS provides a method that uses each Gaussian anchor to generate $k$ Gaussian primitives through MLPs to achieve a compact representation. In Voxel-GS, we adopted the same framework but quantized the position and features of the anchors in Scaffold-GS to generate integer-type Gaussian point clouds, making them compatible with traditional point cloud codecs. The position of the anchors $P$ of Scaffold-GS is quantized and the duplicates of the anchors are removed to reduce redundancy and irregularity, which precisely meets the requirements of the point cloud codecs (\ie, G-PCC). Assuming that the voxel resolution of the anchor position is $Q_P$, we modified the anchor position function in Scaffold-GS to
\begin{equation}
    P_Q = \operatorname{STE\_ROUND}(P*Q_P)/Q_P,
\label{EQ1}
\end{equation}
where $\operatorname{STE\_ROUND}$ is a rounding operation with Straight-Through Estimator~\cite{bengio2013estimating}, which can back propagating the gradient
for training.
For the features that Scaffold-GS needed for representation, we made a similar modification. Specifically, the \texttt{offsets} $O$ with dimensions $3*k$ are quantized with $Q_O$, the \texttt{anchor features} $A$ with dimensions $m$ are quantized with $Q_A$, and the \texttt{scaling factors} $S$ with dimensions $6$ are quantized with $Q_S$. Note that the \texttt{rotation} and the \texttt{opacity} are not required as saved information in Scaffold-GS, because they are derived from the \texttt{anchor features} by tiny MLPs. All required features are warpped into a $\operatorname{STE\_ROUND}$ function. At the end of the reconstruction, the quantized values of those features,
\begin{equation}
    F_Q = \operatorname{STE\_ROUND}(F*Q_F), F\in\{P,O,A,S\}
\label{EQ2}
\end{equation}
are saved as integers, and thus generate a Gaussian point cloud with $3*k+m+6$ dimensions of attributes with $P_Q$ as the coordinate.
 
The factors affecting the size of the Gaussian point cloud includes the number of anchors (\ie, the number of points in the point cloud), attribute dimensions and the value ranges of attributes. Fewer anchors can also accelerate the training process during reconstruction. Thus, we introduce anchor pruning before adaptive control in Voxel-GS, that is, the anchor will be pruned when all $k$ derived Gaussian primitives become invisible. The voxel resolution $Q_P$ affects the number of anchors, and the quantization parameters of the features directly control the value range of attributes. The $m$ and $k$ affect attribute dimensions; however, they also significantly change the number of anchors due to the anchor pruning. This complex factor is beyond the scope of this study. Therefore, $Q_P, Q_O, Q_A, Q_S$ will be considered as hyperparameters and studied in the experiments, and $m$ and $k$ are determined  following HAC++\cite{HAC_Plus}.

\textbf{Rate Proxy}
Because traditional point cloud codecs are non-differentiable, we propose a rate proxy to estimate the bitrate and constrain features to perform rate-distortion optimization during reconstruction. Run-Length Coding (RLC)~\cite{Run_length} is widely used in the entropy coder of traditional point cloud codecs, which is an efficient method, especially when there are large numbers of consecutive identical data. We propose using a Laplace distribution to estimate its probabilistic model,
\begin{equation}
    q(x) = \int_{x-0.5}^{x+0.5}\mathcal{L}_{\mu, \sigma}(x)\,dx,
    \label{EM}
\end{equation}
where the means $\mu$ and the standard deviation $\sigma$ are drawn from all the $x$ to be encoded. The actual bitrate of run-length coding is approximately proportional to the cross-entropy $R\approx\alpha H(p, q)$, where $p$ is the true distribution of quantized features $F_Q$, and $\alpha$ represents the proportion by which run-length coding outperforms the proxy with the assumed Laplace distribution. The loss term of the rate proxy is
\begin{equation}
\ell_{R} = \sum_{F\in \{O,A,S\}} \mathbb{E}_{F_Q}(-\log q(F_Q)).
\end{equation}
Compared to introducing complex deep entropy models, the rate proxy does not require any parameters of the neural network, resulting in a model size that is significantly smaller than that of end-to-end compression methods. This is particularly advantageous in small-scale scenario compression. Note that we have not imposed any constraints on $P_Q$, which will be coded by Octree. Its bitrate is typically proportional to the resolution of the voxel and remains stable during training.

\textbf{Reconstruction and Compression}
We follow the same dataset configuration as the Scaffold-GS. For each scene, a total of 30,000 steps is trained, and anchor pruning and adaptive control are performed from 1,500 to 15,000 steps. The quantization in Eqn.~\eqref{EQ1} and the rate proxy are disabled before 15000 steps to help Voxel-GS generate a stable representation. After 15,000 steps, the number of anchors remains unchanged and the quantization and rate proxy is enabled. Therefore, the final loss function is
\begin{equation}
    \ell= \lambda_1\ell_{L1} + \lambda_2(1-\ell_\text{SSIM}) + \lambda_3\ell_{R}.
\end{equation}
After the reconstruction process, the quantized anchor position $P_Q$ is losslessly compressed by the octree branch of G-PCC (\ie, tmc13v23); the quantized offset $O_Q$, anchor features $A_Q$, and scaling factors $S_Q$ are losslessly compressed by RLC directly (bypass mode) under the reconstructed geometry in Morton order. The tiny MLPs of naive Scaffold-GS are also saved.

\Section{Experimental Results}
\textbf{Datasets and Metrics}
We follow HAC++~\cite{HAC_Plus} to perform evaluations on multiple datasets, including Synthetic-NeRF~\cite{mildenhall2021nerf}, Mip-NeRF360~\cite{barron2022mip}, Tank\&Temples~\cite{knapitsch2017tanks}, and DeepBlending~\cite{hedman2018deep}. PSNR, SSIM, and LPIPS are adopted to evaluate the rendering quality, alongside the SIZE of the bitstream to assess the compression efficiency.

\textbf{Model Parameters and Implementation}
Anchor features dimensions are set to $m=50$ and $k=10$. The voxel resolution $Q_P$ is set to $1024$ for Synthetic-NeRF and $200$ for the other three datasets. The quantization scales $Q_O,Q_A,Q_S$ are set to $1, 1, 8$ for all data, respectively. The weights in loss are set to $\lambda_1=0.2$, $\lambda_2=0.8$, and $\lambda_3=0.0001$\footnote{To balance the differences in the ranges of $\ell_{R}$ (100$\sim$1000) and $\ell_{L1},\ell_\text{SSIM}$ (0$\sim$1).}. Other hyperparameters are kept the same as in Scaffold-GS. The Octree and run-length coding are implemented by C++ backend. The experiments are conducted on a single NVIDIA 4090 GPU with an Intel i9-13900K CPU.

\begin{figure}[h]
    \centering
    \includegraphics[width=\linewidth]{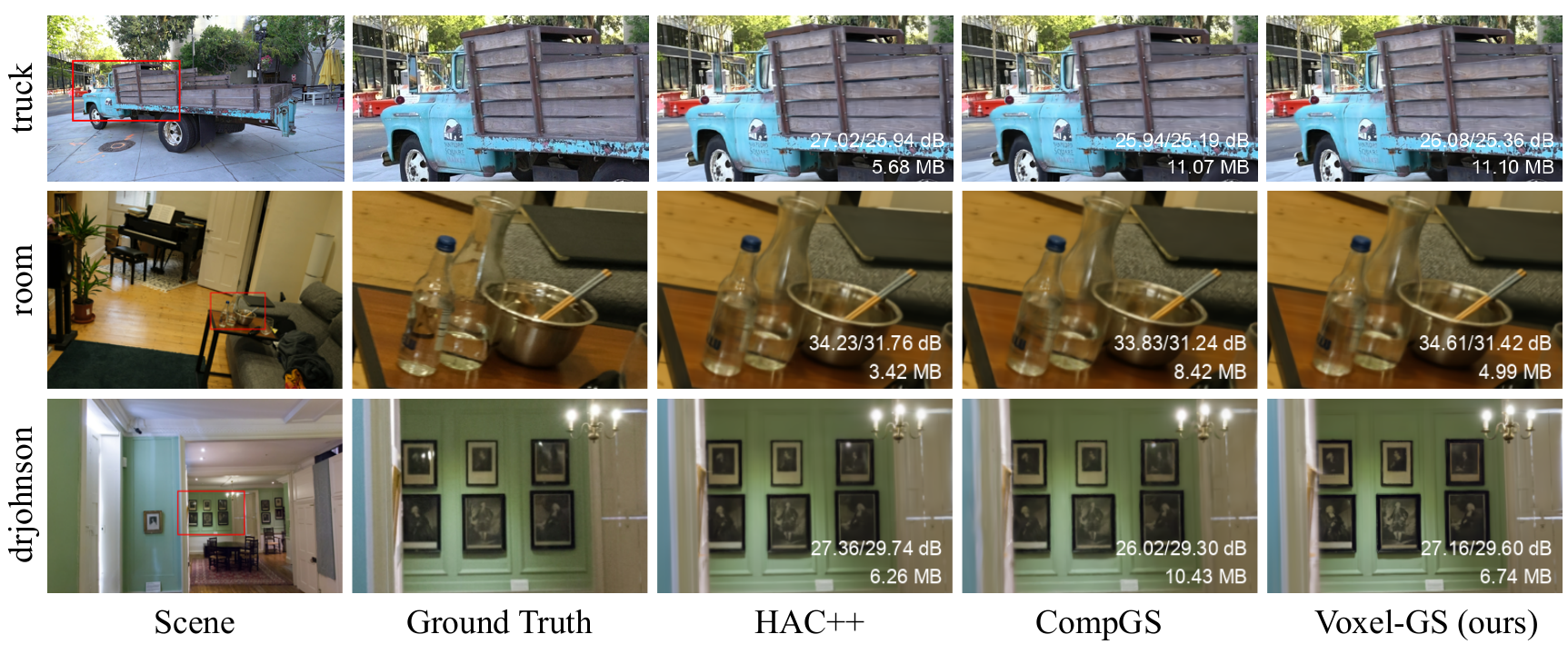}
    \caption{Qualitative comparison. PSNR of this view/all views and size are labeled.}
    \label{Quan_vis}
\end{figure}

\begin{table}[t] {\fontsize{9.51}{12}\selectfont
    \centering
    \setlength\tabcolsep{1.pt}
    \caption{Quantitative results of anchors, A-3DGS and I-3DGS compression methods}\label{table1}
    \scalebox{0.9}{
    \begin{tabular}{c|cccc|cccc|cccc|cccc}
        \toprule
        {\textbf{Datasets}} & \multicolumn{4}{c|}{\textbf{Synthetic-NeRF~\cite{mildenhall2021nerf}}} & \multicolumn{4}{c|}{\textbf{Mip-NeRF360~\cite{barron2022mip}}} & \multicolumn{4}{c|}{\textbf{Tank\&Temples~\cite{knapitsch2017tanks}}} & \multicolumn{4}{c}{\textbf{DeepBlending~\cite{hedman2018deep}}}  \\
        {\textbf{methods}} & psnr$\uparrow$    & ssim$\uparrow$   & lpips$\downarrow$ & size$\downarrow$   & psnr$\uparrow$   & ssim$\uparrow$   & lpips$\downarrow$ & size$\downarrow$   & psnr$\uparrow$   & ssim$\uparrow$   & lpips$\downarrow$ & size$\downarrow$ & psnr$\uparrow$    & ssim$\uparrow$   & lpips$\downarrow$ & size$\downarrow$   \\
        \midrule
3DGS\cite{kerbl20233d}  & 33.80 & 0.970 & 0.031 & 68.46 & 27.46 & 0.812 & 0.222 & 750.9 & 23.69 & 0.844 & 0.178 & 431.0 & 29.42 & 0.899 & 0.247 & 663.9 \\
ScaffoldGS~\cite{lu2024scaffold} & 33.41 & 0.966 & 0.035 & 19.36 & 27.50  & 0.806 & 0.252 & 253.9 & 23.96 & 0.853 & 0.177 & 86.50  & 30.21 & 0.906 & 0.254 & 66.00  \\
       \midrule
HAC\cite{chen2024hac}               & 33.24 & 0.967 & 0.037 & 1.18  & 27.53 & 0.807 & 0.238 & 15.26 & 24.04 & 0.846 & 0.187 & 8.10  & 29.98 & 0.902 & 0.269 & 4.35  \\
CompGS\cite{liu2024compgs}            & -     & -     & -     & -     & 26.37 & 0.778 & 0.276 & 8.83  & 23.11 & 0.815 & 0.236 & 5.89  & 29.30 & 0.895 & 0.293 & 6.03  \\
ContextGS\cite{wang2024contextgs}         & 32.79 & 0.965 & 0.040 & 1.01  & 27.62 & 0.808 & 0.237 & 12.68 & 24.20 & 0.852 & 0.184 & 7.05  & 30.11 & 0.907 & 0.265 & 3.45  \\
HAC++\cite{HAC_Plus}            & 33.03 & 0.966 & 0.039 & 0.88  & 27.60  & 0.803 & 0.253 & 8.34  & 24.22 & 0.849 & 0.190  & 5.18  & 30.16 & 0.907 & 0.266 & 2.91  \\
       \midrule
LightGS\cite{LightGaussian}     & 32.73 & 0.965 & 0.037 & 7.84  & 27.00  & 0.799 & 0.249 & 44.54 & 22.83 & 0.822 & 0.242 & 22.43 & 27.01 & 0.872 & 0.308 & 33.94 \\
Comp3D\cite{niedermayr2024compressed}      & 32.94 & 0.967 & 0.033 & 3.68  & 26.98 & 0.801 & 0.238 & 28.80 & 23.32 & 0.832 & 0.194 & 17.28 & 29.38 & 0.898 & 0.253 & 25.30 \\
FCGS\cite{chen2024fast}             & -     & -     & -     & -     & 27.05 & 0.798 & 0.237 & 34.64 & 23.48 & 0.832 & 0.193 & 17.89 & -     & -     & -     & -     \\
HybridGS\cite{yanghybridgs}          & -     & -     & -     & -     & 25.97 & 0.766 & 0.272 & 21.73 & 23.12 & 0.820 & 0.236 & 11.10 & 29.05 & 0.885 & 0.250 & 16.35 \\
Ours & 32.75 & 0.963 & 0.043 & 0.77 & 26.87 & 0.784 & 0.271 & 12.51 & 23.57 & 0.829 & 0.219 & 7.77 & 30.19 & 0.909 & 0.262 & 6.72\\
    \bottomrule
    \end{tabular}}
    \begin{tablenotes}
        \footnotesize
        \item[] {The size is measured in MB, and psnr is in dB.}
    \end{tablenotes}
    }
\end{table}

\begin{table}[t]
\centering
{
    \setlength\tabcolsep{2.5pt}
    \renewcommand{\arraystretch}{0.9} 
    \caption{Complexity comparison averaged on Tank\&Temples dataset}\label{table2}
\begin{tabular}{c|ccc|c|c|c}
\toprule
Methods    & Train (min) & Enc. (s) & Dec. (s) & FPS   & GPU Mem. & GS number \\ \midrule
Scaffold-GS~\cite{lu2024scaffold} & 13.32       & -        & -        & 112.1 & 7 GB             & 517839    \\
CompGS~\cite{liu2024compgs}     & 46.60       & 6.27     & 4.46     & 99.8  & 13 GB            & 244370    \\
HAC++~\cite{HAC_Plus}      & 24.67       & 8.13     & 13.33    & 105.6 & 10 GB            & 286650    \\
Ours       & 14.01        & 1.28     & 1.47     & 111.8 & 6 GB            & 312238
\\
\bottomrule
\end{tabular}
    }
\end{table}

\textbf{Results}
The visual comparison is illustrated in Fig.~\ref{Quan_vis}, where Voxel-GS shows exceptional visual quality. The quantitative comparison of multiple datasets is shown in Table 1. Among the I-3DGS compression methods, our approach demonstrates the best performance. Compared to the latest HybridGS, our method achieves an average improvement of 0.85 dB while reducing the bitrate by 45\%. Averaged across three large-scale datasets, Voxel-GS (26.88 dB, 17.92 MB) is comparable to CompGS (26.26 dB, 17.47 MB). Furthermore, our method (28.34 dB, 6.94 MB) is very close to HAC (28.70 dB, 7.22 MB) on average across four datasets. Notably, our method outperforms ContextGS in smaller scenes (\ie, Synthetic-NeRF), as the parameterless proxy offers advantages in limited bitstream budgets compared to the deep entropy model. While there is still a slight gap between Voxel-GS and the joint learning approach HAC++ in some datasets, we believe the benefits of I-3DGS compression methods, particularly in terms of flexibility, model size, and speed are more significant. 
\begin{figure}[t]
    \centering
    \hspace{-3.5mm}
    \includegraphics[width=1.02\linewidth]{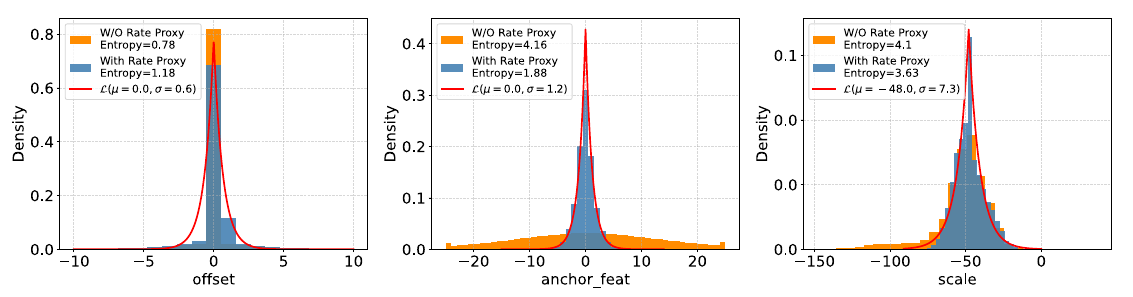}
    \caption{Example histograms of \texttt{offset}, \texttt{anchor feature}, and \texttt{scaling factor} with the estimated Laplacian distribution (red curve) from the “chair” scene.}
    \label{hist}
\end{figure}
\begin{figure}[t]
    \begin{center}
    \begin{tabular}{cc}
    \includegraphics[width=0.45\linewidth]{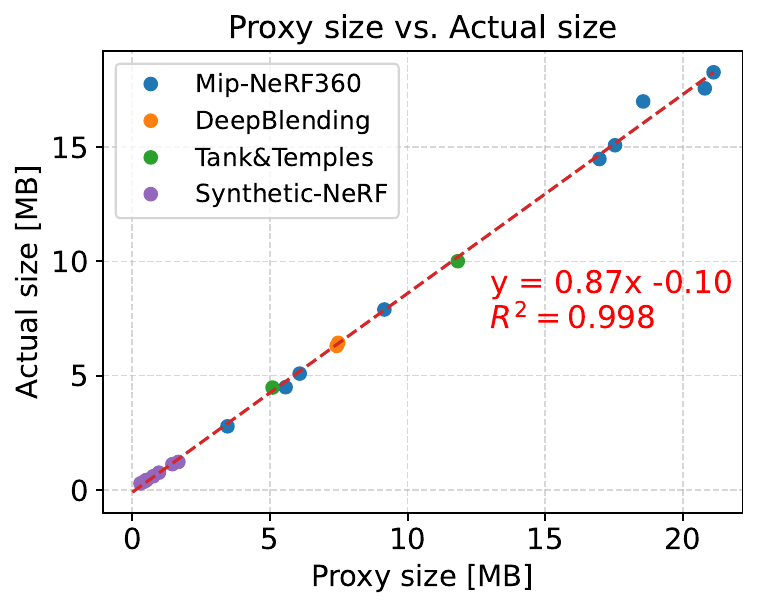} &
    \includegraphics[width=0.505\linewidth]{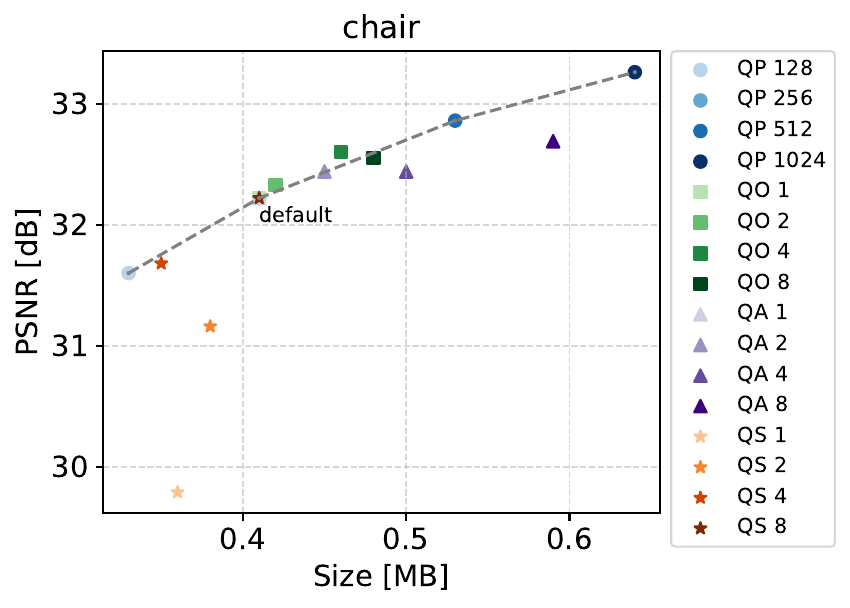} \\
    {\small (a)} & {\small (b)}
    \end{tabular}
    \end{center}
    \caption{\label{proxy_qs}%
    (a) Bitrate of the rate proxy estimation
    and the actual run-length coding. (b) Rate-distortion performance on ``chair'' with different quantization parameters. The default value is $Q_P=256, Q_O=1, Q_A=1, \text{and }Q_S=8$.}
    \end{figure}
\textbf{Complexity}
The complexity comparison tested on the Tank\&Temples dataset is presented in Table~\ref{table2}. In comparison to other methods, Voxel-GS does not rely on complex entropy models during the reconstruction process, enabling it to maintain slightly more Gaussian primitives while still being about 2$\sim$3 times faster than joint-learning methods in training and utilizing almost half the GPU memory. Furthermore, the traditional coder exhibits less complexity than the deep entropy model, allowing us to save 85\% of the encoding and decoding time compared to HAC++. When compressing and decompressing a large number of scenes, this time savings becomes particularly significant, making our approach more applicable.

\textbf{Rate Proxy Effectiveness} To validate the proposed rate proxy, we removed it and performed an ablation study. The results of the ``chair" scene in the Synthetic-NeRF are shown in Fig.~\ref{hist}, and similar results were observed on other datasets. It can be observed that after adding the rate proxy, the features \texttt{anchor feature} and \texttt{scaling factor} are significantly more concentrated around a Laplace distribution, with the entropy reduced by 33\%. For the \texttt{offset}, its distribution seems unexpectedly more dispersed. One possible explanation is that, under the original Scaffold-GS, many offsets are not activated and remain at zero. After adding the constraint, more derived Gaussians are activated to maintain the rendering quality. Overall, the proxy constraint reduced the size of the bitstream from the original 1.59 MB of quantized Scaffold-GS to 0.91 MB, and maintained the rendering quality almost unchanged (34.15 dB to 34.14 dB). This clearly demonstrates that the Laplace rate proxy can effectively generate high-fidelity, more compact Gaussian point cloud representations.  

\textbf{Rate Proxy Accuracy} The rate size of the features (\ie, $O,A,S$) of the rate proxy estimation and run-length coding (RLC) are shown in Fig.~\ref{proxy_qs} (a) by datasets. The slope of the fitted line indicates that $\alpha=0.87$ after Eqn.~\eqref{EM}. This means that RLC is saving approximately 13\% of the bitrate compared to Laplace distribution, indicating that RLC is more efficient. The actual coding bitrate is closely aligned with the estimated bitrate, with a correlation coefficient of 0.998. This indicates that the proposed rate proxy can accurately estimate the rate of run-length coding, providing a reliable basis for the rate-distortion optimization during reconstruction.

\begin{table}[t]
    \centering
    \setlength\tabcolsep{2.2pt}
    \renewcommand{\arraystretch}{0.9} 
    \caption{Bit allocation comparison on large scene and small scene}\label{table3}
\begin{tabular}{c|c|cccccc|cc}
\toprule
Scene                                      & Method & P      & O       & A       & S       & MLP    & Other  & Size (MB) & PSNR \\ \midrule
\multicolumn{1}{c|}{\multirow{2}{*}{train}} & HAC++  & 3.08\% & 25.97\% & 45.67\% & 17.29\% & 4.34\%  & 3.64\% & 7.61  & 22.61    \\
\multicolumn{1}{c|}{}                       & Ours   & 5.71\% & 22.02\% & 56.50\% & 14.62\% & 1.14\%  & 0.00\% & 4.44  & 21.79    \\ \midrule
\multirow{2}{*}{ficus}   & HAC++  & 6.86\% & 10.23\% & 32.28\% & 12.94\% & 32.68\% & 5.03\% & 0.64  & 34.53      \\
 & Ours   & 5.62\% & 13.21\% & 58.18\% & 13.35\% & 9.65\%  & 0.00\% & 0.53  & 34.60
 \\\bottomrule

\end{tabular}
\begin{tablenotes}
    \footnotesize
    \item[] {P: \texttt{anchor positions}, O: \texttt{offsets} A: \texttt{anchor features} and S: \texttt{scaling factors}.}
\end{tablenotes}
\end{table}

\textbf{Parameters Selection}
We analyze the impact of quantization parameters by adjusting one of them from the default point. The default point is set to $Q_P=256, Q_O=1, Q_A=1, \text{and }Q_S=8$. The R-D result on the ``chair'' scene is shown in Fig.~\ref{Quan_vis} (b). Among them, the effect of adjusting $Q_P$ is the most significant and stable (represented by the black dashed line), so it can be used as a regular parameter for multiple rate coding and rate control. In contrast, adjusting $Q_S$ leads to a severe drop in PSNR; therefore, we recommend setting it above 8. The effects of adjusting $Q_O$ and $Q_A$ are not significant and are typically recommended as auxiliary fine-tuning parameters after determining $Q_P$, and the recommended value is 1.

\textbf{Bit Allocation} 
The comparison of bit allocation with HAC++ is presented in Table~\ref{table3}. The table illustrates that the bit ratios for most components in our method exhibit a consistent trend compared to HAC++ excepted for MLP (network). In larger scenes, such as ``train", the network occupies a smaller proportion of the total size, giving HAC++ an advantage over our approach. Conversely, in smaller scenes, such as ``ficus", the limited bitrate budget results in a significant increase in the proportion of the network, which may become a bottleneck for joint learning methods. In contrast, Voxel-GS does not incorporate a deep entropy model and maintains an MLP size of only 0.07 MB, which is advantageous for compressing small scenes.

\Section{Conclusion}
This paper presents Voxel-GS for generating and compressing Gaussian splatting format point clouds. The method introduces a rate proxy based on the Laplace distribution to guide Scaffold-GS in generating compact, integer-type Gaussian point clouds. Additionally, octree and run-length coding are employed for lossless compression of the quantized GS point clouds. The Voxel-GS is comparable in performance to the most advanced methods and has faster training and coding speeds. At the same time, it is also aligned with standardization requirements. We believe that the proposed method contributes to the standardization of Gaussian point cloud compression.

\Section{Appendix}
\begin{table}[h]
\renewcommand{\arraystretch}{0.9} 
\centering{\small
\setlength\tabcolsep{2.2pt}
    \caption{Detailed results of all scenes of Voxel-GS. The gray row is the average.}\label{table4}
\begin{tabular}{ccccc|ccccc}
\toprule
Scene & PSNR$\uparrow$    & SSIM$\uparrow$   & LPIPS$\downarrow$ & SIZE$\downarrow$ & Scene  & PSNR$\uparrow$    & SSIM$\uparrow$   & LPIPS$\downarrow$ & SIZE$\downarrow$ \\ \midrule
chair     & 34.14 & 0.981 & 0.018 & 0.91  & bicycle   & 24.42 & 0.699 & 0.317 & 19.38 \\
drums     & 26.20 & 0.950 & 0.046 & 1.09  & bonsai    & 31.12 & 0.934 & 0.207 & 7.91  \\
ficus     & 34.60 & 0.984 & 0.015 & 0.53  & counter   & 27.86 & 0.884 & 0.245 & 3.38  \\
hotdog    & 36.89 & 0.981 & 0.030 & 0.51  & flowers   & 20.56 & 0.520 & 0.429 & 14.63 \\
lego      & 34.91 & 0.978 & 0.023 & 1.47  & garden    & 26.74 & 0.825 & 0.175 & 20.63 \\
materials & 29.45 & 0.951 & 0.060 & 0.51  & kitchen   & 30.30 & 0.916 & 0.146 & 6.12  \\
mic       & 35.16 & 0.988 & 0.012 & 0.39  & room      & 31.42 & 0.919 & 0.219 & 4.99  \\
ship      & 30.63 & 0.893 & 0.143 & 0.76  & stump     & 26.42 & 0.745 & 0.301 & 17.47 \\
          &       &       &       &       & treehill  & 22.95 & 0.617 & 0.397 & 18.07 \\
\rowcolor{gray!10}
Sync.     & 32.75 & 0.963 & 0.043 & 0.77  & Mip360.   & 26.87 & 0.784 & 0.271 & 12.51 \\
\midrule
train     & 21.79 & 0.789 & 0.259 & 4.44  & drjohnson & 29.60 & 0.906 & 0.261 & 6.74  \\
truck     & 25.36 & 0.868 & 0.180 & 11.10 & playroom  & 30.78 & 0.911 & 0.263 & 6.71  \\
\rowcolor{gray!10}
T2T.      & 23.57 & 0.829 & 0.219 & 7.77  & DB.       & 30.19 & 0.909 & 0.262 & 6.72   \\  
\bottomrule
\end{tabular}
}
\end{table}
\Section{References}
\bibliographystyle{IEEEtran}
\bibliography{refs}

\end{document}